Title: Isolation of Pu-isotopes from environmental samples
using ion chromatography for accelerator mass spectrometry and alpha spectrometry

Corresponding Author: Physicist Elena Chamizo,

Corresponding Author's Institution: Centro Nacional de Aceleradores

First Author: Elena Chamizo

Order of Authors: Elena Chamizo; María del Carmen Jiménez-Ramos, PHD Student; Lukas Wacker, doctor; Ignacio Vioque, Doctor; Ana Calleja, PHD student; Rafael García-Tenorio, professor; Manuel García-León, professor

Abstract: A radiochemical method for the isolation of plutonium isotopes from environmental samples, based on the use of specific chromatography resins for actinides (TEVA®, Eichrom Industries), has been set up in our laboratory and optimised for their posterior determination by alpha spectrometry (AS) or accelerator mass spectrometry (AMS). The proposed radiochemical method has replaced in our lab a well-established one based on the use of a relatively un-specific anion-exchange resin (AG® 1X8, Biorad), because it is clearly less time consuming, reduces the amounts and molarities of acid wastes produced, and reproducibly gives high radiochemical yields.
In order to check the reliability of the proposed radiochemical method for the determination of plutonium isotopes in different environmental matrixes, twin aliquots of a set of samples were prepared with TEVA and with AG 1X8 resins and measured by AS. Some samples prepared with TEVA resins were measured as well by AMS. As it is shown in the text, there is a comfortable agreement between AS and AMS, which adequately validates the method.

* Cover Letter

# Isolation of Pu-isotopes from environmental samples using ion chromatography for accelerator mass spectrometry and alpha spectrometry

E. Chamizo [a]*, M.C. Jiménez-Ramos [b], L. Wacker [c], I. Vioque [b], A. Calleja [b], M. García-León [a,b], R. García-Tenorio [b]

[a] *Centro Nacional de Aceleradores (CNA), Avda. Thomas A. Edison , 41092, Sevilla, Spain.*

[b] *Applied Nuclear Physics Group, University of Sevilla, P.O Box 1065, 41080, Sevilla, Spain.*

[c] *Institute of Particle Physics, ETH Hönggerberg, CH-8093 Zürich, Switzerland.*

* Manuscript

# Isolation of Pu-isotopes from environmental samples using ion chromatography for accelerator mass spectrometry and alpha spectrometry

## Abstract

A radiochemical method for the isolation of plutonium isotopes from environmental samples, based on the use of specific chromatography resins for actinides (TEVA®, Eichrom Industries), has been set up in our laboratory and optimised for their posterior determination by alpha spectrometry (AS) or accelerator mass spectrometry (AMS). The proposed radiochemical method has replaced in our lab a well-established one based on the use of a relatively un-specific anion-exchange resin (AG® 1X8, Biorad), because it is clearly less time consuming, reduces the amounts and molarities of acid wastes produced, and reproducibly gives high radiochemical yields.

In order to check the reliability of the proposed radiochemical method for the determination of plutonium isotopes in different environmental matrixes, twin aliquots of a set of samples were prepared with TEVA and with AG 1X8 resins and measured by AS. Some samples prepared with TEVA resins were measured as well by AMS. As it is shown in the text, there is a comfortable agreement between AS and AMS, which adequately validates the method.

## 1.- Introduction

Plutonium is an anthropogenic element that has found its way into the environment since the beginning of the nuclear age in 1945. Among the different actinides released due to nuclear activities, Pu deserves a special attention from the point of view of radioecology and toxicology. The main source of Pu to the environment were the atmospheric nuclear weapons tests carried out mainly in the 1950's and the 1960's, which introduced 0.3 TBq of Pu in a global scale [1]. Additional amounts of this element have been released at a local scale as a consequence of civil nuclear reactor accidents (as the Chernobyl accident), releases from nuclear reprocessing facilities (as Sellafield, UK; La Hague, France; Mayak PA, Russia) and nuclear weapons accidents (Palomares, Spain; Thule, Greenland).

Among the four major plutonium isotopes ($^{241}Pu$, $^{238}Pu$, $^{239}Pu$ and $^{240}Pu$), three are alpha-emitters: $^{238}Pu$ ($T_{1/2}$=87.7 y, $E_\alpha$=5,499 MeV (72%)), $^{240}Pu$ ($T_{1/2}$=6564 y, $E_\alpha$=5,168 (76%)) and $^{239}Pu$ ($T_{1/2}$=24110 y, $E_\alpha$=5,157 (73,3%)). Conventionally, they are determined in environmental samples by alpha spectrometry (AS). However, the two most abundant plutonium isotopes, $^{240}Pu$ and $^{239}Pu$, cannot be quantified independently in low-level samples due to their very similar alpha-energy emissions ($\Delta E$=11 keV). AS usually informs about the joined activity of $^{239}Pu$ and $^{240}Pu$. In many cases, this is enough for a simple monitoring or surveillance program.

Nevertheless, the study of the ratio of $^{240}Pu$ and $^{239}Pu$ offers unambiguously information about the origin of the plutonium. The $^{240}Pu/^{239}Pu$ atomic ratio ranges from 0.03 to 0.06 for weapon-grade plutonium, and reaches a value of 0.4 for reactor-grade plutonium. This ratio can be quantified by mass spectrometry techniques (MS): either thermal ionisation mass spectrometry (TIMS) [2], inductively coupled mass spectrometry (ICP-MS) [3], or accelerator mass spectrometry (AMS) [4]. Among them, AMS offers very low detection limits (~1 µBq for both $^{239}Pu$ and $^{240}Pu$), making it possible to study these isotopes in any environmental compartment. On the other hand, limited information on the Pu source term can be gained by measuring the $^{238}Pu/^{239+240}Pu$ activity ratio by AS.

We routinely perform in our laboratory the determination of $^{238}Pu$ and $^{239+240}Pu$ in a wide variety of environmental matrixes by AS. Recently, we have gained as well experience in the field of the determination of $^{239}Pu$ and $^{240}Pu$ by AMS [5]. In both cases, it is necessary to apply a radiochemical procedure to isolate the plutonium fraction and adapt it to optimum physicochemical conditions for the final measurement: a) the electro-deposition of plutonium onto stainless steel disks for its determination by AS or b) the dispersion of the plutonium in a metal/metal oxide matrix for its measurement by AMS.

In this paper, we describe a radiochemical method, based on the use of the extraction chromatography resin TEVA (Eichrom Industries), for the isolation of the plutonium from different environmental matrixes. The following points have been studied: recovery yields, alpha-peak resolution in AS, and the magnitude of the U, Th and Dy decontamination factors as potential interferences for AMS or AS. The necessary validation was performed by applying the TEVA method and the conventional one, based on the AG 1-X8 resin (Biorad), to different samples, and comparing the obtained results by AS. Also some independent aliquots were prepared with the TEVA method and measured by AMS. That way, the validity of the recently adapted TEVA method for both AS and AMS has been checked.

A description of the TEVA method is given in Section 2, its main characteristics and advantages are highlighted in Section 3 and, finally, the results obtained in the validations exercises are shown and discussed in Section 4.

## 2.- Experimental

### 2.1.- Standards, resins, equipments and samples

The samples have to be spiked with a known amount of $^{242}$Pu both for AS and for AMS measurements. This spike is necessary in both techniques for the determination of plutonium concentrations by the isotope-dilution method: $^{239+240}$Pu activity in AS, or $^{239}$Pu and $^{240}$Pu activities in AMS. In AS, this spike can be used as well to estimate the chemical recovery yields of the procedure. The standard solution of $^{242}$Pu was supplied by NPL, England, and was diluted to obtain stock solutions with a concentration of 20.9 mBq/ml (144 pg/ml) in a 2M $HNO_3$ matrix.

The extraction chromatography resin used was TEVA (Eichrom Industries), which has been proved to be a good option for the separation of tetravalent actinides ($Pu^{4+}$, $Np^{4+}$ and $Th^{4+}$) in 2-4 M nitric acid from $Am^{3+}$ and $U^{6+}$ [6]. Ready-to-use mini-columns (diameter: 7 mm; length: 8 cm) packed with 2 ml of resin (100-150 µm mesh) were used for these analyses.

The AS system used for the measurements was an alpha spectrometer *Alpha Analyst* (Canberra) with eight independent chambers that can work in parallel. Each chamber was equipped with a 450 $mm^2$ passivated implanted planar silicon detector (*PIPS*, 18 keV of nominal resolution). The measurements were performed by placing the samples at a distance of 1.5 mm from the detector. In these conditions a counting efficiency of 34% was achieved [7].

.

The AMS measurements were performed with the ETHZ/PSI compact AMS system *Tandy* adapted for the measurement of actinides. Briefly, the plutonium was extracted from the $Cs^+$ sputtering ion-source as $PuO^-$, stripped to $Pu^{3+}$ with an Ar stripper gas at a terminal voltage of 330 kV, and counted with a gas ionisation detector with 1.3 MeV of energy. This detector features with a 3x3 $mm^2$ silicon nitride entrance window, which is 40 nm thick for optimal separation of the plutonium ions from the $2^+$ molecular fragments with the same M/q ratio [4].

The ICP-MS determinations, carried out to study the magnitude of the U, Th and Dy decontamination factors, were done with an Agilent-7500C mass spectrometer.

The analysed samples were the followings: a) artificially traced water, b) ashes of wood, c) superficial soils collected in Palomares (Spain) and d) superficial soils with fallout plutonium. Some of these samples were used either in the decontamination factors study or in the validation exercises.

## 2.2.- Proposed TEVA radiochemical method

***Sample pre-treatment*** *( 20 h)*

A known amount of $^{242}Pu$ (1 ml of the 20.9 mBq/ml solution for AS, and 20 µl of the same solution for AMS) was added to each sample (water, soils, wood) as a spike, together with some ml of diluted HCl for its homogenisation.

The solid samples were dried at 70ºC for 2 hours. Afterwards, they were ashed in a muffle furnace to remove the organic matter in two sequential steps: firstly at 150ºC during 2 hours, and finally at 550ºC during 6 hours. The water sample was directly evaporated to dryness.

Once cooled, the residues were placed in a teflon beaker and 20 ml of 8M $HNO_3$ and 2 ml of $H_20_2$ were added. After heating at 90ºC for 1 hour, 1 ml of concentrated HF was added in order to enhance the quantitative extraction of the refractory plutonium that may be present, particularly in the analysed soils from Palomares. The mixture was then heated for 5 hours, adding 8M $HNO_3$ to keep the volume at 20 ml. The supernatant was thereby isolated by centrifugation (4000 rpm, 10 minutes). The solid residue was discarded.

The supernatant was then evaporated to dryness and the resulting residue was dissolved with 1 ml 8M $HNO_3$ and 10 ml of distilled water, giving a 0.7M $HNO_3$ solution. The plutonium was then subjected to a redox cycle: reduction to $Pu^{3+}$ with $Fe^{2+}$, adding 0.1 g of $FeSO_4$, and oxidation to $Pu^{4+}$, adding 8 ml of 8M $HNO_3$ and 0.2 g of $NaNO_2$. At these conditions, the plutonium forms a negatively charged nitrate complex, $Pu(NO_3)_6^{2-}$, which is retained in the column. The obtained solution (about 19 ml in a 3 M $HNO_3$ medium) was finally conditioned to a 0.5M $Al(NO_3)_3$ medium by adding 3 g of $Al(NO_3)_3$-$9H_20$, to complex the oxalates that reduce the retention of the plutonium in the column [6]. The sample was then loaded into the extraction chromatography resin TEVA.

***Separation by TEVA resins (4 h)***

After the conditioning of the commercially pre-packed 2 ml TEVA resin with 10 ml of 3M $HNO_3$, the sample solution was loaded onto the column. At these conditions, the Pu is stick to the resin, and most of the U and matrix elements are eluted (see posterior discussion).

The resin was then washed successively with three different acid solutions: firstly with 10 ml of 3M $HNO_3$, secondly with 40ml of 2M $HNO_3$ to assure the removal remaining traces of U and, finally, with 15 ml of 6M HCl to elute the Th.

In a last step the Pu was eluted with 20 ml 0.5 M HCl. The Pu fraction was then ready for the preparation of the Pu source in the form required for each technique.

***Preparation of the Pu source***

The final plutonium fraction has to be adapted to an optimum physicochemical medium for each technique. In the case of AS, the plutonium in our laboratory is conventionally deposited onto a stainless-steel disk, following a well-established electrodeposition method [8]. For AMS the plutonium, as plutonium oxide, needs to be dispersed in a metallic matrix for its measurement in a $Cs^+$ sputtering ion-source.

To prepare the AMS target, 0.5 mg of $Fe^{3+}$ (as $Fe(NO_3)_3$) was added to the final plutonium fraction from the TEVA resin (20 ml of 0.5 M HCl) as a carrier. This solution was evaporated to dryness (5 h). The residue was baked at 800ºC in a muffle furnace for 3 hours to convert the plutonium and iron to oxides ($Pu_3O_8$, $Fe_2O_3$). Finally, 2 mg of pure aluminium powder were mixed with the oxides and pressed into a 1 mm diameter aluminium cathode. That way, the sample was ready to be measured by AMS [4, 9].

A complete flow-diagram of the analytical TEVA procedure proposed in this paper for Pu determination in environmental samples is shown in Fig. 1.

### 2.3. - Conventional method

This radiochemical procedure is fully described in [7]. Briefly, it consists of the following steps:

Step 1.- The samples are ashed (in order to remove the organic matter) and then wet-oxidised to put Pu-isotopes and other elements in dissolution.
Step 2.- After filtration, the dissolved sample is passed through an AG® 1-X8 anion exchange resin, from where after several elutions (in order to remove interfering elements), the Pu-isotopes are stripped in a very purified solution.
Step 3.- The Pu-isotopes are finally electrodeposited onto stainless-steel planchets. The Pu sources so obtained are then ready for measurement.

## 3.- Results and Discussion

***U, Th and Dy decontamination factors***

In order to check the reliability of the TEVA method for both AS and AMS, the quality of the plutonium fraction from the TEVA resin has been studied. In both cases, the achievement of a purified plutonium fraction is important: in AMS, the $PuO^-$ current extracted in the sputtering ion source improves with the concentration of Pu in the target; in AS, the resolution of the energy spectra is closely related to the thickness of the electrodeposits. Moreover, special attention must be paid to the decontamination factors achieved for some trace elements that may be potential plutonium interferences in AS ($^{228}$Th and $^{241}$Am) or AMS ($^{238}$U and $^{160}$Dy).

As long as alpha spectrometry is concerned, it is important to remove traces of $^{241}$Am ($T_{1/2}$=432.2 y, Eα=5.486 MeV (85%)) and $^{228}$Th ($T_{1/2}$=1.91 y, Eα=5.420 MeV (73%)) in the final plutonium fraction, as their main alpha-emissions interfere with the ones of $^{238}$Pu ($T_{1/2}$=87.7 y, Eα=5.499 MeV (72%)) in the energy spectra.

In the case of the compact AMS system Tandy, $^{238}$U may reach the detector even when the machine is tuned for $^{239}$Pu, due to straggling on the residual gas in the vacuum system or due to instrumental instabilities. The mass suppression factor observed for $^{238}U$ on its neighbouring mass $^{239}Pu$ is about $10^{-7}$ [4]. These $^{238}U$ ions cannot be discriminated in the final detection system in any case, so the massive presence of $^{238}U$ in the target increases the detection limit for $^{239}Pu$. On the other hand, working with $Pu^{3+}$ after the stripping process, the molecular fragments in the $2^+$ charge state, with the same M/q ratio as the plutonium ions, reach the detector. Although they have a different energy and can be discriminated with an appropriate detector, pile-up events may generate spurious plutonium counts (Fig. 2.). The most critical case is that of $^{160}Dy^{2+}$, which has exactly the same M/q ratio as $^{240}Pu^{3+}$. Therefore, the achievement of good Dy and U decontamination factors is essential in low-energy AMS.

The decontamination factors obtained for $^{238}U$, $^{160}Dy$ and $^{232}Th$ (or $^{228}Th$) with the proposed TEVA procedure were studied for soil and water matrixes by ICP-MS. These decontamination factors were calculated by comparing the concentration of these elements in the initial loaded solution to their concentration in the final plutonium fraction. The obtained results are displayed in Fig.3. They are higher than 100 for $^{238}U$, at least 100 for $^{232}Th$ and of the order of $10^4$ for $^{160}Dy$. Regarding the measurement of $^{239}Pu$ by low-energy AMS, the achieved $^{238}U$ decontamination factors make the study of $^{239}Pu$ at fg levels, when the $^{238}U$ concentration in the sample is about 1 μg, feasible. An additional step for uranium decontamination becomes redundant. In the case of AS, the Am decontamination factor can be extrapolated from the results got for Dy, as they may be in the same oxidation step ($3^+$) during the elution process and may behave in the same way. This fact, together with the high decontamination factors obtained for Th isotopes, point out a reliable determination of $^{238}Pu$ in environmental samples. To check this, the proposed radiochemical procedure was applied to measure the $^{238}Pu$ concentration in one inter-comparison sample (CIEMAT 2005, [10]). The obtained result ($A(^{238}Pu)=0.106 \pm 0.012$ mBq/g) matches with the expected one ($A(^{238}Pu)=0.111 \pm 0.020$ mBq/g).

***Spectra resolution in Alpha Spectrometry***

Another important point is the quality of the deposits obtained in AS with the new TEVA procedure. In Fig. 4. the obtained spectra for the twin aliquots of Soil-2 that were prepared with the TEVA procedure and with the conventional one are displayed. The samples were measured during the same time and at the same distance from the detector (1.5 mm). The improvement in the resolution of the peaks and the cleanliness of the spectra obtained with the TEVA procedure are remarkable. In Table 1, the FWHM (in keV) of the $^{242}$Pu alpha emissions are shown for both procedures. On average, the resolution improves with the TEVA method in a factor of 1.6.

### *Radiochemical Yields*

The TEVA radiochemical method highlight also by the high and stable radiochemical yields that can be obtained. In Table 2, the obtained recovery factors for $^{242}$Pu with the two methods by AS are displayed. In each case, the radiochemical yield obtained by applying the TEVA method was quantitative higher that the ones obtained by the AG 1X8 (on average, 68% for the TEVA, and 52% for the AG 1X8 method).

### *Validation Exercises*

In order to check the feasibility of the proposed TEVA method for both AS and AMS, a set of environmental samples were measured with both techniques. In the case of AS, the TEVA method and the AG 1X8 one were applied to twin aliquots of solid and water samples. Fractions of 5 g of the solid samples and of 0.5 l of the water sample were used for the determinations. For the AMS measurements, only the TEVA procedure was applied for some of the soil matrixes. Aliquots of 1 g were used in this case.

The obtained results by AS are shown in Table 3, and the ones got by AMS are displayed in Table 4. For the sake of comparison, the results got by AS for the samples measured by AMS are given as well in Table 4. In all the cases, there is an agreement between the results got with the two radiochemical methods and the two techniques. Moreover, taking into account the wide range of $^{239+240}$Pu activity concentrations studied, it can be concluded that the proposed radiochemical procedure makes possible the quantitative determination of the plutonium alpha-emitters with either AMS or AS.

Regarding the plutonium isotopic composition, both techniques produce reliable results, as the $^{238}$Pu/$^{239+240}$Pu activity ratios got by AS, and the $^{240}$Pu/$^{239}$Pu atomic ratios obtained by AMS, agree with the expected ones attending to the origin of the samples. The samples Soil-1 and Soil-5 correspond to superficial soils collected in the region of Palomares (Spain). Therefore, they may have traces of the weapon-grade plutonium that was released over this region in 1966 due to a nuclear accident [10]. The measured $^{240}Pu/^{239}Pu$ atomic ratio for Soil-5 by AMS was 0.11% [5], and the $^{238}Pu/^{239+240}Pu$ activity ratio got by AS for Soil-1 was close to 2% [11]. These ratios show the contribution of the weapon-grade plutonium released in the nuclear accident. However, the $^{240}Pu/^{239}Pu$ atomic ratios obtained by AMS for Soil-2 and Soil-3 were close to 18%, and the $^{238}Pu/^{239+240}Pu$ activity ratio were about 3%, which are the expected numbers for fallout plutonium [5, 7]. The latter samples were collected in an environment that hasn't been affected by local plutonium releases.

## 5.- Conclusions

From the results we can conclude that the proposed radiochemical method, based on the use of the extraction chromatography resins TEVA (Eichrom Industries), is fully applicable for the accurate determination of the plutonium in different environmental matrixes by both AMS and AS. This TEVA method highlights for the high selectivity in the isolation of the plutonium, its high recovery yields, and the good decontamination factors obtained. At the same time, compared to conventional anion-exchange resins, the amounts of strong acids and the working time have been reduced.

## 6.- Acknowledgements

This work has been financed through the projects FIS2004-0495 of the Spanish Ministry of Science and Education, and RNM419 of the *Junta de Andalucía*. The authors are indebted to the group of Particle Physics of the ETH/PSI, Zürich, for offering the compact AMS system Tandy for the measurements.

## 7.- References

**Fig. 1.**

Flow-diagram of the analytical TEVA procedure proposed in this paper for the plutonium-isotopes determination in environmental samples. The plutonium fraction from the TEVA resin was adapted for either AS or AMS.

**Fig. 2.**

Spectra obtained for $^{240}Pu$ during an AMS measurement with the ETH/PSI compact AMS system Tandy. The $^{80}Se^{1+}$ and $^{160}Dy^{2+}$ molecular fragments reach the detector because they have exactly the same M/q ratio as the $^{240}Pu^{3+}$. A miniaturised gas ionisation detector, provided with a 3x3 $mm^2$ silicon nitride window 40 nm thick, was used for these measurements [4].

**Fig. 3** .

Decontamination factors for $^{238}U$, $^{232}Th$ and $^{160}Dy$ in the plutonium fraction purified with the TEVA resins for two different soils and artificially traced water, according to the procedure explained in this work. Only $^{238}U$ and $^{232}Th$ were added to the water sample.

**Fig. 4.**

Superimposed plutonium spectra obtained by AS for the twin aliquots of Soil-2 that were prepared by the proposed TEVA method and by the conventional one (AG 1X8). The resolution of the peaks and the cleanliness of the spectra improve in the case of the TEVA procedure.

**Table 1**

Full Width at Half Maximum (FWHM) in keV, obtained for the $^{242}Pu$ alpha emission in the electroplated sources obtained with the TEVA and the conventional method for twin aliquots of different samples

**Table 2**

Plutonium recovery yields obtained by AS in the analysis of aliquots of several environmental samples by the proposed TEVA method and by the AG 1X8 one. They are systematically higher in the first case.

**Table 3**

Results for the $^{239+240}Pu$ activity concentration for twin aliquots of different environmental samples measured by applying the conventional and the proposed TEVA method by AS.

**Table 4**

Comparison between the $^{239+240}Pu$ activity concentration measured by AS and AMS for different environmental samples.

**Table 1**

| Sample | AG 1X8 | TEVA |
|---|---|---|
| Soil-1 | 115 | 68 |
| Soil-2 | 60 | 36 |
| Soil-3 | 70 | 44 |
| Water | 50 | 35 |

**Table 2**

| Sample | Pu recovery yield (%) TEVA method | Pu recovery yield (%) AG 1-X8 method |
|---|---|---|
| Water | 76 | 61 |
| Soil-1 | 74 | 54 |
| Soil-2 | 62 | 55 |
| Soil-3 | 68 | 51 |
| Ashes | 60 | 39 |

**Table 3**

| | **AS** | |
|---|---|---|
| **Sample** | **$^{239+240}Pu$ (Bq/kg) TEVA method** | **$^{239+240}Pu$ (Bq/kg) AG 1X8 method** |
| **Water** | 35.0 ± 2.2 Bq/m$^3$ | 36.0 ± 2.1 Bq/m$^3$ |
| **Ashes** | 0.08 ± 0.01 Bq/kg | 0.08 ± 0.01 Bq/kg |
| **Soil-1 (Palomares)** | 49.4 ± 1.0 Bq/kg | 46.5 ± 2.3 Bq/kg |
| **Soil-2 (Fallout)** | 2.40 ± 0.15 Bq/kg | 2.37 ± 0.09 Bq/kg |
| **Soil-3 (Fallout)** | 1.48 ± 0.08 Bq/kg | 1.44 ± 0.07 Bq/kg |

**Table 4**

| | AS | | AMS |
|---|---|---|---|
| **Sample** | **$^{239+240}Pu$ (Bq/kg) TEVA method** | **$^{239+240}Pu$ (Bq/kg) AG 1X8 method** | **$^{239+240}Pu$ (Bq/kg) TEVA method** |
| **Soil-2 (Fallout)** | 2.40 ± 0.15 | 2.37 ± 0.09 | 2.65 ± 0.14 |
| **Soil-3 (Fallout)** | 1.48 ± 0.08 | 1.44 ± 0.07 | 1.53 ± 0.09 |
| **Soil-4 (Fallout)** | | 1.55 ± 0.06 | 1.50 ± 0.05 |
| **Soil-5 (Palomares)** | | 0.06 ± 0.01 | 0.049 ± 0.005 |

Figure 1
Click here to download high resolution image

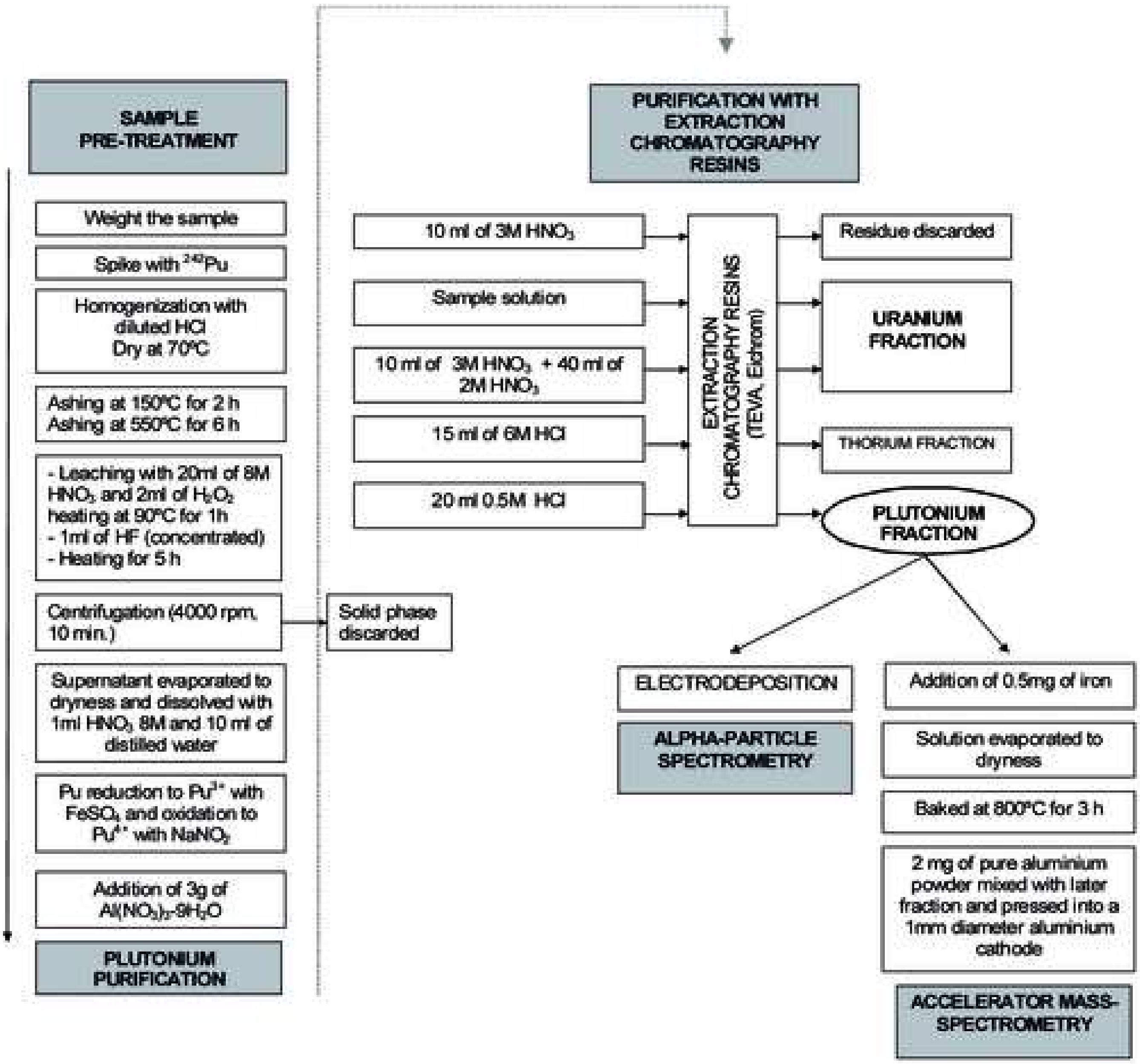

Figure 2

Click here to download high resolution image

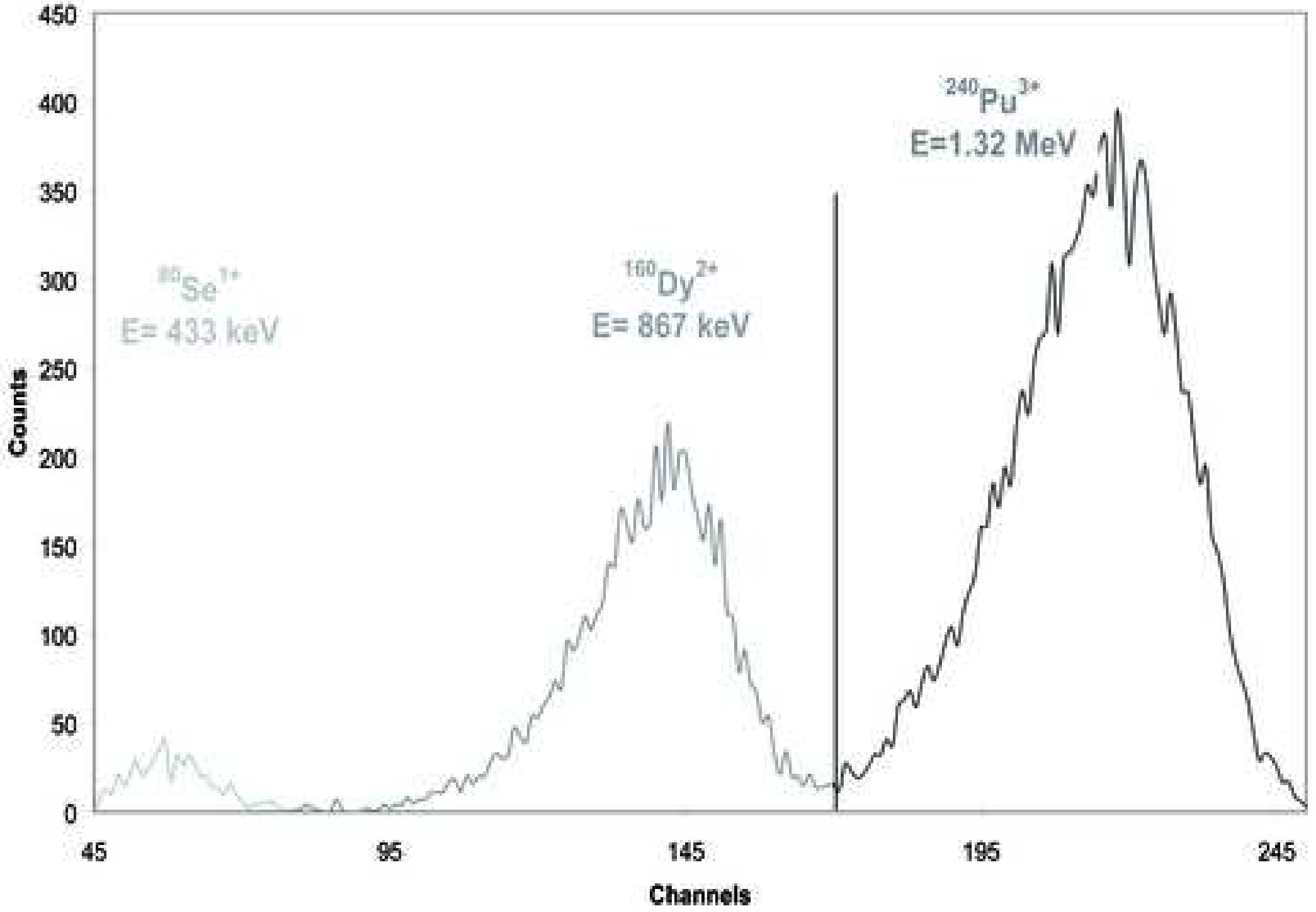

**Figure 3**
**Click here to download high resolution image**

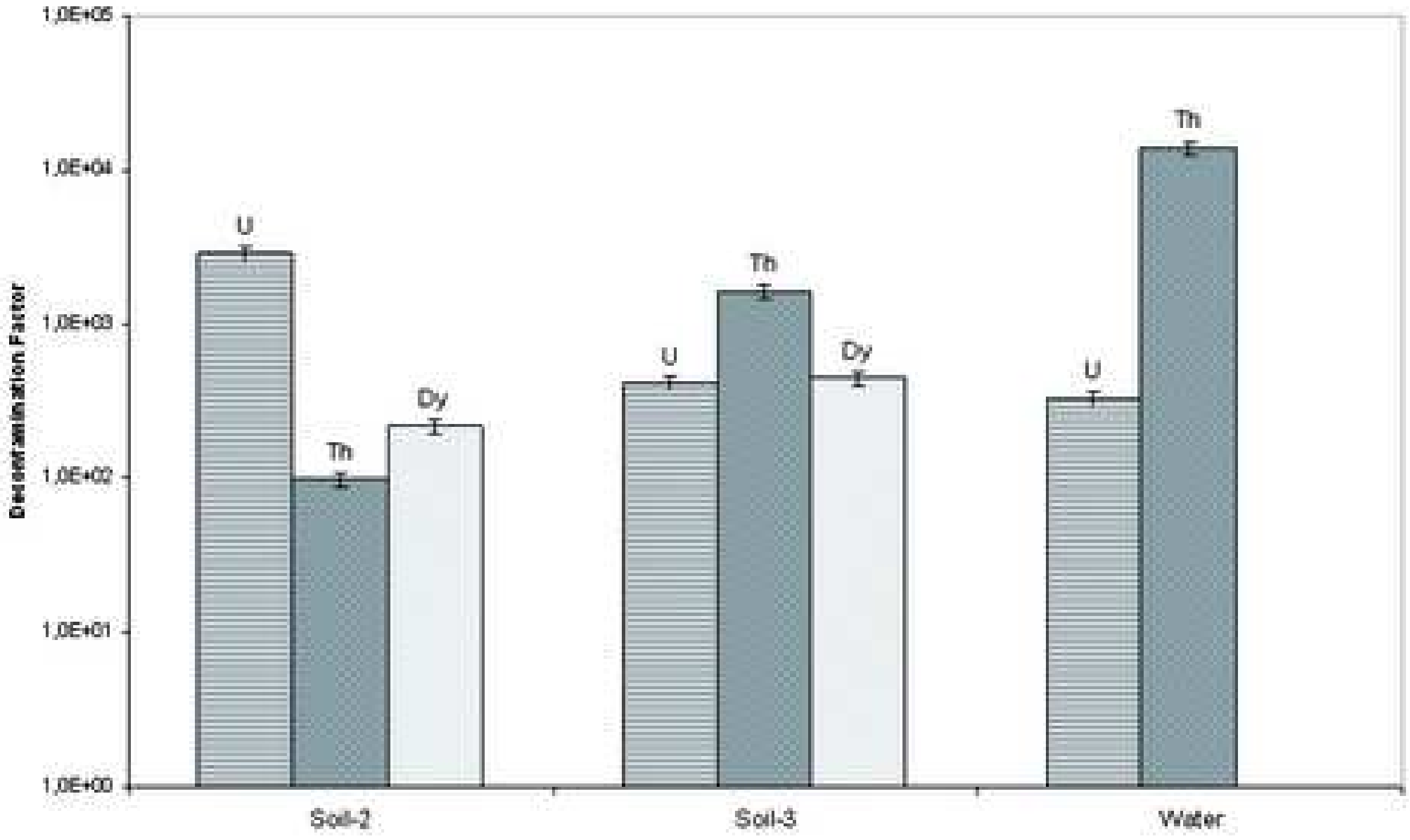

**Figure 4**
**Click here to download high resolution image**

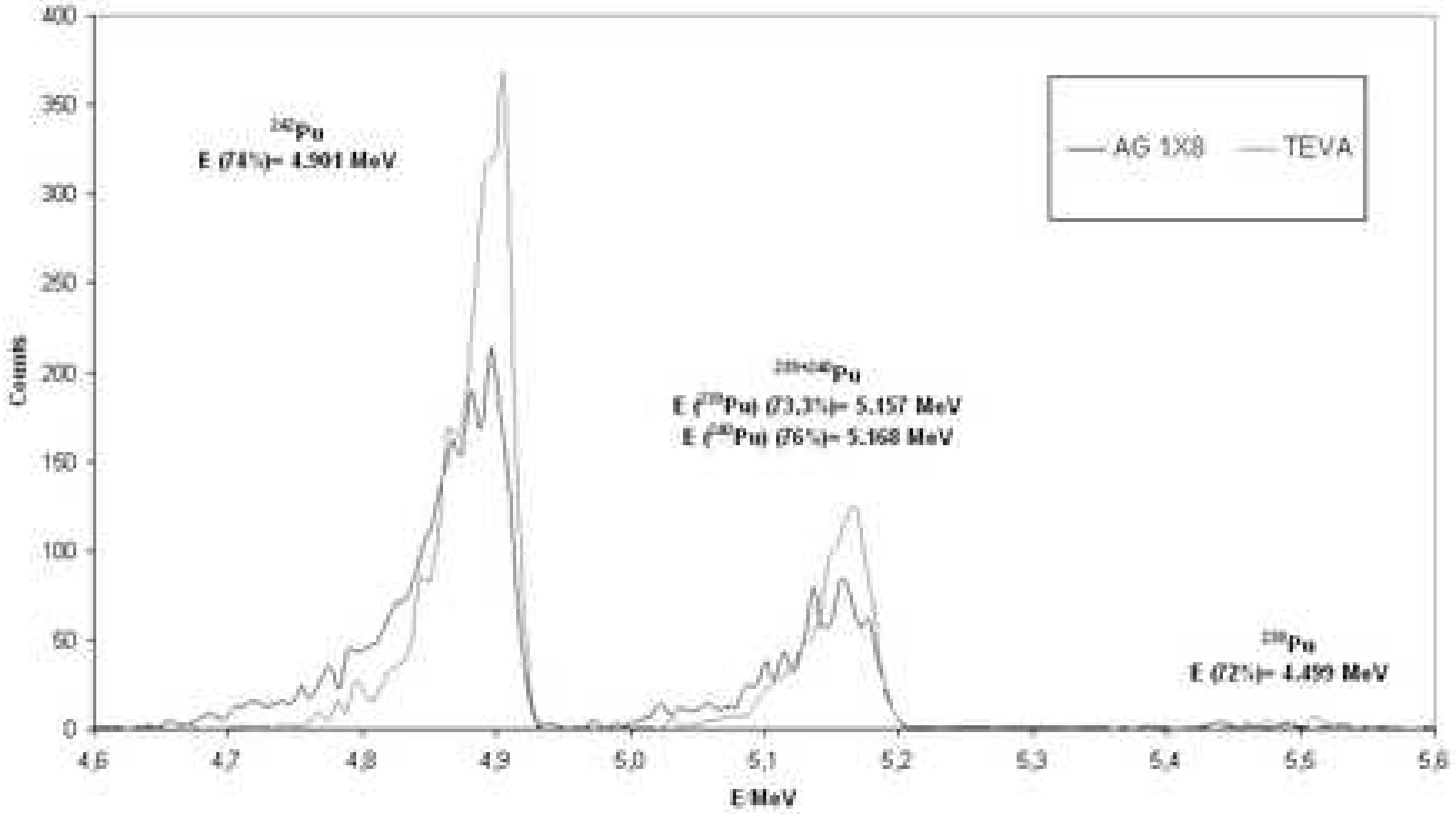